\documentclass[aps,prl,twocolumn,superscriptaddress,showpacs]{revtex4}
\usepackage{epsfig}
\usepackage{latexsym}

\begin{document}     

\title{Magneto-transport in copper-doped noncentrosymmetric BiTeI} 

\author{Chang-Ran Wang}
\affiliation{Institute of Physics, Academia Sinica, Nankang, Taipei 11529, Taiwan}
\author{Jen-Chuan Tung}
\affiliation{Graduate Institute of Applied Physics, National Chengchi University, Taipei 11605, Taiwan}
\author{R. Sankar}
\affiliation{Center for Condensed Matter Sciences, National Taiwan University, Taipei 10617, Taiwan}
\author{Chia-Tso Hsieh}
\affiliation{Institute of Physics, Academia Sinica, Nankang, Taipei 11529, Taiwan}
\author{Yung-Yu Chien}
\affiliation{Institute of Physics, Academia Sinica, Nankang, Taipei 11529, Taiwan}
\author{Guang-Yu Guo}
\affiliation{Graduate Institute of Applied Physics, National Chengchi University, Taipei 11605, Taiwan}
\affiliation{Department of Physics and Center for Theoretical Sciences, National Taiwan University, Taipei 10617, Taiwan}
\author{F. C. Chou}
\affiliation{Center for Condensed Matter Sciences, National Taiwan University, Taipei 10617, Taiwan}
\author{Wei-Li Lee}\email{wlee@phys.sinica.edu.tw}
\affiliation{Institute of Physics, Academia Sinica, Nankang, Taipei 11529, Taiwan}
\date{\today}      

\begin{abstract}

BiTeI exhibits large Rashba spin splitting due to its noncentrosymmetric crystal structure. The study of chemical doping effect is important in order to either tune the Fermi level or refine the crystal quality. Here, we report the magneto-transport measurement in high quality BiTeI single crystals with different copper dopings. We found that a small amount of copper doping improves the crystal quality significantly, which is supported by the transport data showing higher Hall mobility and larger amplitude in Shubnikov-de Haas oscillation at low temperature.  Two distinct frequencies in Shubnikov-de Haas oscillation were observed giving extremal Fermi surface areas of $A_{S} = 9.1 \times 10^{12}$ cm$^{-2}$ and $A_{L} = 3.47\times 10^{14}$ cm$^{-2}$ with corresponding cyclotron masses $m_S^*$ = 0.0353 $m_e$ and $m_L^*$ = 0.178 $m_e$, respectively. Those results are further compared with relativistic band structure calculations using three reported Te and I refined or calculated positions. Our analysis infers the crucial role of Bi-Te bond length in the observed large bulk Rashba-type spin splitting effect in BiTeI.      
                   
\end{abstract}
\maketitle
BiTeI emerges as an intriguing material that shows a large Rashba effect \cite{ishizaka,eremeev,sakano} and a possible topological phase transition under pressure \cite{byan}. Its crystal structure comprises alternating layers of bismuth (Bi), tellurium (Te) and iodine (I) each with trigonal planar lattice as illustrated in Fig. 1(a). It was proposed \cite{shevelkov} to constitute a semi-ionic structure along the stacking direction, where $\rm (BiTe)^+$ layer is positively charged and (Bi-I) layer is ionic. Angle-resolved photo-emission spectroscopy experiments (ARPES) \cite{ishizaka,crepaldi} have revealed evidence for the giant Rashba spin splitting, and its bulk nature was further confirmed by bulk-sensitive optical spectroscopy \cite{lee} and soft x-ray ARPES \cite{landolt}. When comparing to band structure calculation, the Te and I coordinations turn out to be crucial parameters that can result in dramatic difference in the calculated band property. There are three different Te and I coordinations reported in the literature: coordination A 
with Te(2/3,1/3,0.6928) and I(1/3,2/3,0.2510) from the refinement analysis of X-ray experiment \cite{shevelkov}, 
as well as coordination B with Te(2/3,1/3,0.7111) and I(1/3,2/3,0.2609)\cite{kulbachinskii}, and coordination C 
with Te(2/3,1/3,0.7482) and I(1/3,2/3,0.3076) \cite{bahramy}, from two different theoretical structural 
determinations using the same band structure method. Regardless of the small variation, only coordination C with a shortest Bi-Te bond length ($d_{\rm Bi-Te}$ = 3.05 $\rm\AA$) gives rise to a giant Rashba spin-splitting in the bulk band with a Rashba parameter $\alpha_R \cong$ 5.4 eV$\rm\AA$ according to our calculations, which may infer a close connection between $d_{\rm Bi-Te}$ and its Rashba effect. 

In this paper, we show magneto-transport measurement results on high quality Cu$_x$BiTeI single crystals with copper (Cu) doping $x$ up to 0.2. Comparing to earlier works on Shubnikov-de Haas (SdH) oscillations \cite{martin,bell}, the SdH oscillation in our crystals exhibits two distinct frequencies derived from a large Fermi surface (LFS) and a small Fermi surface (SFS), which is an order of magnitude larger in amplitude at similar temperatures. The corresponding Fermi surface areas and cyclotron masses can then be unambiguously determined and compared to relativistic band structure calculations using three different atomic coordinations A, B and C with progressive reduction in $d_{\rm Bi-Te(I)}$. Our experimental results, including the angular dependence of SdH frequencies, are in good agreement with the calculation using coordination C and thus provide a strong evidence for the bulk nature of the large Rashba spin-splitting effect. 

\begin{figure}[ht]
\centerline 
{\epsfig{figure=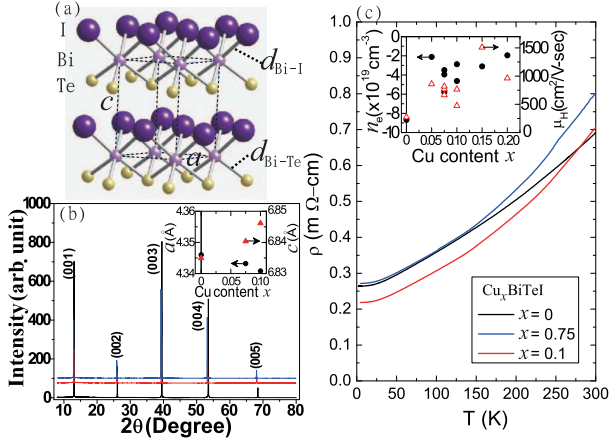,width=8cm,clip=0}}
\caption {\label{xray} (color online) (a) An Illustration of the BiTeI crystal structure. $d_{\rm Bi-Te(I)}$ is the Bi-Te(I) bond length. (b) shows the powder X-ray diffraction pattern of ground Cu$_x$BiTeI single crystals with $x$ = 0, 0.075 and 0.1. The inset figure shows the $x$ dependence on the lattice constants $a$ and $c$. (c) The resistivity $\rho(T)$ for $x$ = 0, 0.075 and 0.1 crystals. The upper inset plots the carrier density $n_e$ (solid circle) and the corresponding Hall mobility $\mu_H$ (open triangle) at 5 K versus Cu content $x$.} 
\end{figure}

Single crystals of BiTeI were grown by direct mixing of pristine elements of Bi, Te and I with additional room temperature agglomeration procedure \cite{sankar}. The powder X-ray diffraction patterns of ground BiTeI crystals with Cu content $x$ = 0, 0.075 and 0.1 are shown in Fig. \ref{xray}(b), where the lattice parameters were determined using space group $\rm\it{P3m1}$ to be $a$ = $b$ = 4.3421 $\rm\AA$ and $c$ = 6.8835 $\rm\AA$ for the pristine sample of $x$ = 0. As $x$ increases, the lattice expands more along the stacking direction as shown in the inset of Fig. \ref{xray}(b). The in-plane resistivity $\rho$ shown in Fig. \ref{xray}(c) exhibits metallic behavior down to 5 K below which it becomes nearly temperature independent. The carrier density $n_e$ and Hall mobility $\mu_H$ were obtained from the Hall effect measurement and shown in the upper inset of Fig. \ref{xray}(c). It is quite evident that the addition of Cu effectively reduces the electron concentration in BiTeI and improves the carrier mobility by nearly an order of magnitude, where SdH oscillation appears with a large amplitude and enables the extraction of the band parameters to be compared with relativistic band calculations.

\begin{figure}[ht]
\centerline 
{\epsfig{figure=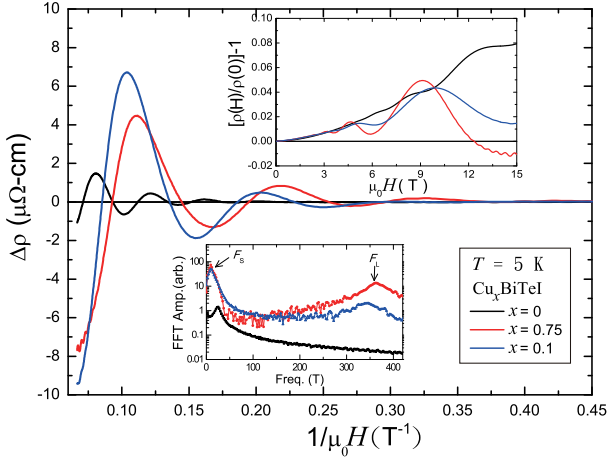,width=8cm,clip=0}}
\caption {\label{sdh} (color online) The oscillatory component of the resistivity $\Delta\rho$ as a function of 1/$\mu_0H$ at 5 K for $x$ = 0, 0.075 and 0.1 crystals. The upper inset shows the magneto-resistance before subtracting the non-oscillatory component of the resistivity. The corresponding FFT spectrum is shown in the lower inset, where two local extremes at $F_S$ and $F_L$ can be clearly identified. } 
\end{figure}  

The upper inset of the Fig.\ref{sdh} shows the magneto-resistance (MR) $[\rho(H)/\rho(0)]-1$ for $x$ = 0, 0.075 and 0.1 as a function of magnetic field up to 15 T at $T$ = 5 K. It exhibits positive MR and starts to show SdH oscillations above 3 T. The pure oscillatory component $\Delta\rho$ in the resistivity \cite{SOM} is extracted and plotted as a function of $1/\mu_0H$ shown in Fig. \ref{sdh}. We remark that the oscillation amplitude is at least 3-fold larger comparing to the undoped for samples with $x$ = 0.075 and 0.1 as shown in red- and blue-lines, respectively. By taking the fast Fourier transformation (FFT) of the MR data, two apparent peaks at $F_S$ and $F_L$ for $x$ = 0.075 and 0.1 were identified in the FFT spectrum shown in the lower inset of Fig. \ref{sdh}, where the suffixes S and L are referred as deriving from a SFS and a LFS, respectively. 

\begin{figure}[ht]
\centerline 
{\epsfig{figure=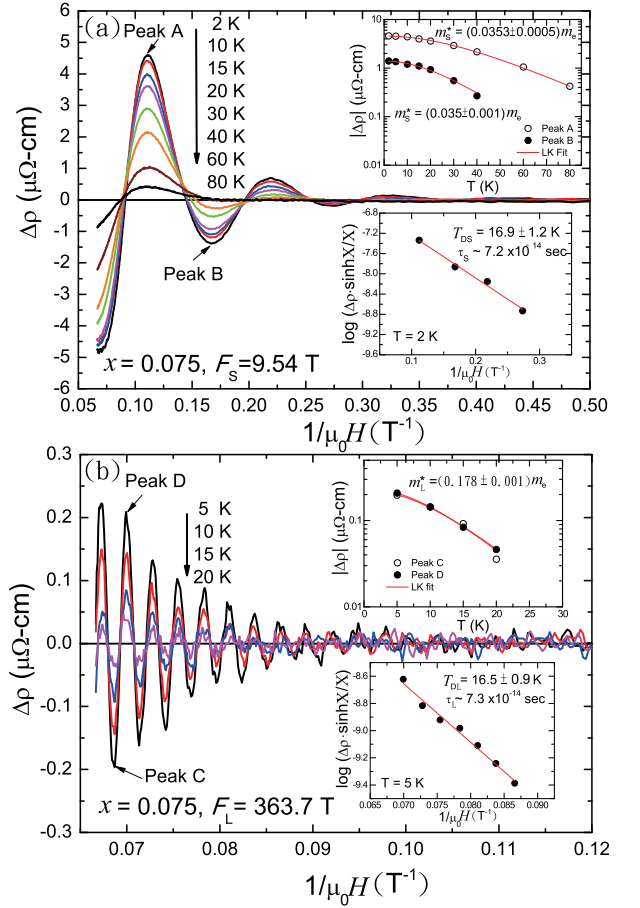,width=8cm,clip=0}}
\caption {\label{mtd} (color online) (a) SdH oscillations of SFS at 8 different temperatures up to 80 K. The temperature dependence of amplitude $|\Delta\rho|$ at Peak A (open circle) and Peak B (solid circle) are shown in the upper inset. The red lines are fitting curves using LK formula for effective cyclotron mass $m^*$ determination. The lower inset shows the linear fit to ${\rm log}(\Delta\rho\cdot {\rm sinh}X/X)$ at local peaks versus 1/$\mu_0H$, where the Dingle temperature $T_D$ can be calculated from the slope. Similarly, (b) shows the SdH oscillations of LFS at 4 different temperatures. The $m^*$ and $T_D$ are determined via LK formula fitting as shown in the upper inset and lower inset, respectively. } 
\end{figure} 

According to Lifshitz-Kosevich (LK) formalism, $\rm\Delta\rho$ can be expressed as 
\begin{equation}
\frac{\Delta\rho(T,B)}{4\rho_0} = {\rm exp}[-X(T_D,B)]\frac{X(T,B)}{{\rm sinh}(X(T,B))},
\label{eq:mynum}
\end{equation}
where $\rho_0$ is the non-oscillatory component of the zero-field resistivity, $X (T,B)\equiv 2\pi^2k_BTm^*/\hbar eB$, $m^*$ is the effective cyclotron mass and $T_D$ is the Dingle temperature. Figure \ref{mtd}(a) and (b) show the oscillatory component $\Delta\rho$ as a function of 1/$\mu_0H$ at different temperatures in $x$ = 0.075 sample arising from SFS and LFS, respectively. In Fig. \ref{mtd}(a), the SdH frequency $F_S$ equals 9.54 T with oscillation amplitude as large as 5 $\rm\mu\Omega cm$ at 2 K, where it remains observable with temperature as high as 80 K. By fitting the temperature dependence of $\Delta\rho$ with LK formula shown in the upper inset in Fig. \ref{mtd}(a), we obtained consistent effective cyclotron masses $m_S^* = 0.0353 \pm 0.0005$ $m_e$ and $0.035 \pm 0.001$ $m_e$ for peak A and peak B locations, respectively, where $m_e$ is the electron rest mass. For $T_{DS}$ determination, we plot ${\rm log}(|\Delta\rho|\cdot {\rm sinh}X/X)$ at local extremes as a function of their corresponding 1/$\mu_0H$ as demonstrated in the lower inset of Fig. \ref{mtd}(a). The data points can be linearly fitted based on LK formula giving a $T_{DS}$ = $16.9 \pm 1.2$ K which corresponds to an electron scattering lifetime $\tau_S\equiv \hbar/2\pi k_BT_{DS} \cong 7.2 \times 10^{-14}$ sec. On the other hand, Fig. \ref{mtd}(b) shows the SdH oscillation arising from LFS giving a $F_L$ = 363.7 T. The effective cyclotron mass $m_L^*$ equals $0.178 \pm 0.001$ $m_e$ from the LK fitting using either peak C or D as shown in the upper inset of Fig. \ref{mtd}(b). The $T_{DL}$ is about $16.5 \pm 0.9$ K giving a $\tau_L \cong 7.3\times 10^{-14}$ sec. The extremal Fermi surface area $A_{e}$ perpendicular to the field direction can be deduced from the SdH oscillation frequency via the Onsager formula $F = \frac{\hbar}{2\pi e}A_e$, which gives $A_{S} = 9.1 \times 10^{12}$ cm$^{-2}$ and $A_{L} = 3.47 \times 10^{14}$ cm$^{-2}$ for SFS and LFS, respectively.

\begin{figure}[ht]
\centerline 
{\epsfig{figure=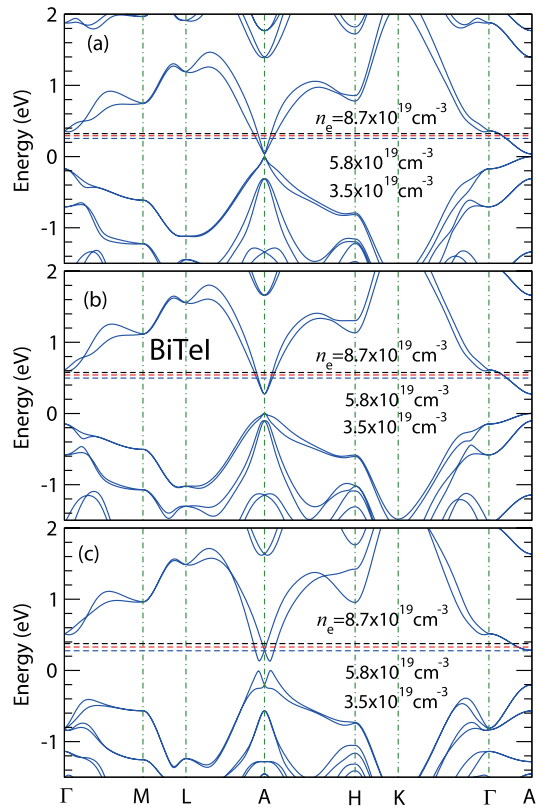,width=7cm,clip=0}}
\caption {\label{band} (color online) Relativistic band structure of BiTeI in the trigonal P3m1 structure using coordination A from Ref. \cite{shevelkov} (a) and as well as the theoretically determined coordination B \cite{kulbachinskii} (b) and coordination C \cite{tung} (c). The top of the valence band is at 0 eV. The lower, middle and upper dashed horizontal lines denote the Fermi level for electron concentrations $n_e$ = 3.5, 5.8 and 8.7 $(\times 10^{19}$ cm$^{-3})$, respectively.}
\end{figure}  

To gain insight into our magneto-transport experiments, we perform 
relativistic band structure calculations for BiTeI within the density functional theory with the generalized
gradient approximation (GGA)\cite{Perdew96} by using the accurate projector augmented-wave (PAW) method, 
as implemented in the VASP package\cite{vasp1,vasp2}. All three coordinations A\cite{shevelkov}, B\cite{kulbachinskii} 
and C\cite{tung} were considered. A large plane-wave cut-off energy of 250 eV was used.
The calculated band structures are shown in Fig. \ref{band} where three horizontal dashed lines in black, 
red and blue denote the Fermi level for electron density $n_e$ = 8.7, 5.8 and 3.5 $\rm(\times 10^{19} cm^{-3})$, 
respectively, which were set based on the experimental electron densities in crystals of $x$ = 0, 0.075 and 0.1. 
We remark that large bulk Rashba spin splitting occurs only for coordination C. Using a rigid band assumption, the Cu doping merely decreases the carrier density of the system and hence shifts the Fermi level lower. We found that experimental values of $F$ and $m^*$ at three different carrier densities of $n_e$ = 3.5, 5.8 and 8.7 ($\times 10^{19}$ cm$^{-3}$) as listed in Table \ref{tab} are much more close to the calculated values using coordination C, which gives the shortest bond length $d_{\rm Bi-Te}$ = 3.05 $\rm\AA$. 

\begin{table}
\caption{Comparison of calculated SdH frequency $\rm\it{F}$(T) and effective mass $m^*(m_e)$ using 
three different coordinations to experimental data. $d_{\rm Bi-Te(I)}$ (\AA) denotes the Bi-Te(I) bond length. 
Calculated Rashba parameter $\alpha_R = 0$ for coordinations A and B. For coordination C, 
$\alpha_R = 5.46$ (5.35) eV\AA$ $ along the AL (AH) direction (see Fig. \ref{band}(c)).} 
\centering 
\begin{tabular}{|lc|rr|rr|rr|} 
\toprule
\multicolumn{2}{|c|}{$n_e (10^{19}$ cm$^{-3})$}&\multicolumn{2}{c|}{3.5}&\multicolumn{2}{c|}{5.8}&\multicolumn{2}{c|}{8.7} \\ [0.5ex]
\colrule
\multicolumn{2}{|c|}{SdH parameters}&\it{F}&$m^*$&\it{F}&$m^*$&\it{F}&$m^*$\\
\multicolumn{2}{|c|}{              }&(T)&($m_e$)&(T)&($m_e$)&(T)&($m_e$)\\
\colrule 
 Coordination A\footnote{Te(2/3,1/3,0.6928), I(1/3,2/3,0.2510)\cite{shevelkov}}&LFS & 175 & 0.203 & 247& 0.247& 317& 0.285 \\ 
 \footnotesize{$d_{\rm Bi-Te(I)}$ = 3.27 (3.04)}&SFS & 72.7 & 0.077 & 99.5& 0.091& 125& 0.104\\
\colrule 
Coordination B\footnote{Te(2/3,1/3,0.7111), I(1/3,2/3,0.2609)\cite{kulbachinskii}} &LFS& 151 & 0.143& 210& 0.173& 268& 0.2 \\
\footnotesize{$d_{\rm Bi-Te(I)}$ = 3.19 (3.08) }&SFS& 104 & 0.088& 140& 0.101& 173& 0.114\\
\colrule 
Coordination C\footnote{Te(2/3,1/3,0.7458), I(1/3,2/3,0.3133)\cite{tung}} &LFS& 319 & 0.167& 397& 0.178& 475& 0.196\\
\footnotesize{$d_{\rm Bi-Te(I)}$ = 3.05 (3.30) }&SFS& 0 & - & 4.4& 0.023& 18.8& 0.047\\
\colrule 
Experimental data &LFS& 339 & 0.186& 364& 0.178& -&- \\
 &SFS& 9.2 & 0.037 & 9.5& 0.035& 24.8& 0.068\\
\botrule 
\end{tabular}
\label{tab}
\end{table}

\begin{figure}[ht]
\centerline 
{\epsfig{figure=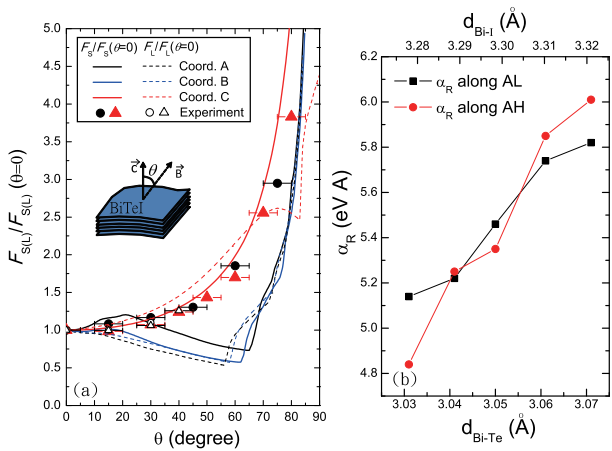,width=8cm,clip=0}}
\caption {\label{angular} (color online) (a) Relative SdH oscillation frequency $F_{S(L)}/F_{S(L)}(\theta = 0)$ as a function of $\theta$ for two $x$ = 0.075 crystals are shown as solid (open) symbols. $\theta$ is defined as the angle between the BiTeI stacking direction $\vec{c}$ and magnetic field direction $\vec{B}$ as illustrated in the inset cartoon. The solid (dashed) lines are the calculated $F_{S(L)}/F_{S(L)}(\theta = 0)$ using coordination A, B and C. The experimental data points agree well with the calculated curves using coordination C within the error. (b) Calculated Rashba parameter $\alpha_R$ versus given Bi-Te bond lengths ($d_{\rm Bi-Te}$) close to that of coordination C.} 
\end{figure} 

We also performed angular dependence of the SdH oscillation for two $x$ = 0.075 crystals in the same batch to further identify the shape of Fermi surface \cite{SOM}. The normalized SdH frequency $F_{S(L)}/F_{S(L)}(\theta = 0)$ as a function of field angle $\theta$ is plotted in Fig. \ref{angular}(a), where the closed (open) symbols and solid (dashed) lines are experimental data and calculated values for SFS (LFS), respectively. The field angle $\theta$ is defined as the angle between the stacking direction $\vec{c}$ and external field as illustrated in the inset cartoon of Fig. \ref{angular}(a). Even though the SdH frequency differs slightly from sample to sample in the same batch, we remark that the normalized SdH frequency $F_{S(L)}/F_{S(L)}(\theta = 0)$ remains to follow well with the calculated band using coordination C as shown in the circles and triangles in Fig. \ref{angular}(a). For more than 4 crystals of $x$ = 0.075 in the same batch we measured, $F_S(\theta = 0)$ values fall in a range of $3.6-9.54$ T with corresponding electron density in a range of $3.5-5.8 \times 10^{19}$ cm$^{-3}$. This somewhat large variation in $F_S(\theta = 0)$ turns out to have no apparent correlation with $n_e$ \cite{SOM} and thus infers a sizable variation of the $\alpha_R$ even in the same batch of Cu$_x$BiTeI crystals.  

The Cu dopings are most likely achieved through the internal redox reaction of the intercalated Cu in the van der Waals gap. While there is no superlattice observed from both powder X-ray and Laue diffraction results up to about 20 \% Cu intercalation per formula unit, the intercalated Cu should distribute randomly or in various domain sizes within the van der Waals gap. Locally, the induced strain by Cu can be relieved by expanding the lattice along the stacking direction (the inset of Fig. \ref{xray}(b)) and also by distorting both the $d_{\rm Bi-Te}$ and $d_{\rm Bi-I}$ in the same manner. It is then reasonable to describe the various degree of residue strain of a layered system with an average $d_{\rm Bi-Te}$ change. By assuming the change of $d_{\rm Bi-I}$ scales linearly with $d_{\rm Bi-Te}$ due to Cu intercalation, the calculated $\alpha_R$ values along AL and AH directions can vary as much as 20 \% (6 - 4.8 eV\AA) while $d_{\rm Bi-Te(I)}$ merely drops by 1 \% (3.07 - 3.03 \AA) as demonstrated in Fig. \ref{angular}(b), which suggests a nontrivial role of average $d_{\rm Bi-Te}$ in the observed Rashba spin splitting effect. In Table \ref{tab}, both the calculated $F$ and $m^*$ values for coordination C drop monotonically with decreasing electron density $n_e$ while the ratio of $A_L/A_S (=F_L/F_S$) increases with descending $n_e$. Apparently, the experimental values in Table \ref{tab} gives a smaller $F_L/F_S$ ratio at similar carrier densities that can not be fully explained by the discrepancy in $n_e$. We, therefore, attribute the likely source of deviation to the difference in average $d_{\rm Bi-Te}$ that dictates $\alpha_R$. Nevertheless, further investigation is keenly required to show how does $d_{\rm Bi-Te(I)}$ in BiTeI influence the charge distribution in $\rm (BiTe)^+$ layer and hence the Rashba spin splitting effect.

In conclusion, Cu$_x$BiTeI is a remarkable system, where we have demonstrated a robust impact of a minor change in atomic coordinations to its band property. The Cu doping not only effectively reduces the electron density but also boosts the carriers' Hall mobility. We observed two distinct frequencies in SdH oscillation, where the corresponding effective cyclotron masses were determined and compared to theoretical band calculations using three different atomic coordinations for Te and I. Our experimental data agree well with the calculation using coordination C with a shortest $d_{\rm Bi-Te}$, which indicates a close connection between $d_{\rm Bi-Te}$ and the giant bulk Rashba spin splitting effect in BiTeI. In principle, $\alpha_R$ can be readily tuned by controlling the BiTeI bond length if applicable. Our finding offers a new possibility for engineering the Rashba spin splitting in a layered and noncentrosymmetric material. \\             


The authors acknowledge the funding support from National 
Science Council in Taiwan. 
W.L.L. acknowledges the funding support from Academia Sinica 2012 career development award in Taiwan.


\clearpage

\begin{thebibliography}{1}
\bibitem{tung} We have also determined the Te and I atomic positions theoretically
by fully relativistic GGA\cite{Perdew96} calculations 
using the PAW method\cite{vasp1,vasp2}. Our theoretical atomic positions
of Te(2/3,1/3,0.7458) and I(1/3,2/3,0.3133) are nearly
identical to that reported in \cite{bahramy} (the bond length differences being within 1.0 \%).
\end{thebibliography}
\end{document}